\documentclass[aps,prd,13,onecolumn,preprintnumbers,floatfix,nofootinbib]{revtex4-1}
\pdfoutput=1
\usepackage{graphicx}
\usepackage{bm}
\usepackage{times}
\usepackage{slashed}
\usepackage{color}
\usepackage{slashed}
\usepackage{amsmath}
\usepackage{amsthm}
\usepackage{subfigure}

\newcommand{\be}{\begin{equation}}
\newcommand{\ee}{\end{equation}}
\newcommand{\bea}{\begin{eqnarray}}
\newcommand{\eea}{\end{eqnarray}}

\begin{document}
\title{Exotic Leptons: Collider and Muon Magnetic Moment Constraints.}
\author{D. Cogollo}
\affiliation{Departamento de Fisica, Universidade Federal de Campina Grande,
Caixa Postal 10071, 58109-970, Campina Grande, PB, Brazil}


\begin{abstract}
In light of the ongoing effort on reducing the theoretical uncertainties and an upcoming experiment concerning muon magnetic moment, we perform a detailed study of an 3-4-1 electroweak gauge extension of the standard model that contains exotic charged leptons in its spectrum. We discuss flavor changing neutral current, collider and electroweak bounds on the model and derive $1\sigma$ limits using current and projected limits on the muon magnetic moment. In summary, we exclude the masses of new gauge bosons that couple to muons and heavy charged leptons up to $700$~GeV. Moreover, we find a projected lower bound on the scale of symmetry breaking to be $2$~TeV.

\end{abstract}

\pacs{12.10.Dm, 12.15.Ff, 12.60.Cn} 
\maketitle

\section{\label{sec:intr}Introduction}

Numerous precision tests of the Standard Model (SM) have taken placed in the last few decades. The results confirmed the SM as the best description of nature we have. Moreover, those tests have also provided stringent constraints on many new physics effects. A clear example is given by the measurements of the anomalous magnetic moment of the electron and the muon, parametrized in terms of $a_e$ and $a_{\mu}$, where recent experiments reached an impressive precision. The $a_{\mu}$ has a relative enhancement of $(m_{\mu}/m_e)^2 \sim 4\times10^4$ compared to the electron anomalous magnetic moment, therefore, $a_{\mu}$ is much better suited to unveil or limit new physics effects, on the other hand, the $a_{\mu}$ allows to test the entire SM since each sector yields a sizeable correction \cite{PDG}. The tau magnetic moment would provide an even better sensitive, but its relatively short lifetime makes a direct measurement impossible, at least at present. Thus, we will focus our study on the muon magnetic moment hereafter. The present difference $\Delta a_{\mu}=a_{\mu}^{exp} - a_{\mu}^{SM}=295 \pm 81 \times 10^{-11}$ yields a $3.6\sigma$ discrepancy \cite{bennet,bennet2}. The large theoretical uncertainties stemming from the the hadronic vacuum polarization and the hadronic contribution to the light-by-light scattering can overshadow the significance of this discrepancy. In the near future important improvements in both the theoretical and experimental situations are expected. Combining the expected progress from the theoretical side, along with the projected experimental sensitivity for the g-2 experiment at Fermilab, the precision will likely reach $\Delta a_{\mu}=295 \pm 34 \times 10^{-11}$, possibly increasing the magnitude of the signal up to $5\sigma$ \cite{carey}. Hence, it is worthwhile to investigate which particle physics models are capable of addressing the muon magnetic moment and also use the current and projected sensitive to derive robust bounds.\\

In this work, we will focus our effort on an electroweak extension of the standard model known as 3-4-1 model. This model extends $SU(2)_L$ SM gauge group to a $SU(4)_L$ one. The motivations for considering such class of models rely on the following:\\
(i) They explain the number of generations by a combination of QCD asymptotic freedom and anomaly cancellation \cite{Foot:1994ym};\\
(ii) They might have plausible dark matter candidates inherited for embedding the 3-3-1 models in the context of the Higgs portal \cite{Mizukoshi:2010ky,deS.Pires:2010fu,Queiroz:2014yna,Queiroz:2014pra,Alvares:2012qv}, $Z^{\prime}$ portal \cite{Profumo:2013sca,Cogollo:2014jia,Alves:2014yha,Alves:2013tqa,Alves:2015pea,
Patra:2015bga,Alves:2015mua}, and dark radiation \cite{Kelso:2013nwa,Kelso:2013paa,Hooper:2011aj,Allahverdi:2014bva,
Queiroz:2014ara,Queiroz:2013lca} capable of addressing interesting indirect detection signals \cite{Hooper:2012sr,Gonzalez-Morales:2014eaa}\\
(iii) They are generally consistent with current collider data \cite{Sanchez:2008qv};\\

The muon magnetic moment has been addressed in the context of 3-4-1 models, but in different versions \cite{Cogollo:2014aka,Cogollo:2014tra}. In those previous studies exotic charged leptons were absent. The presence of exotic charged leptons induce profound changes in the computations and conclusions derived from the muon magnetic moment. Hence, we will restrain our study to a 3-4-1 model which contains heavy charged leptons. Flavor changing neutral current processes and the spontaneous symmetry breaking mechanisms have been fully investigated in such model. However, there is a lack of phenomenological studies, specially related to the muon magnetic moment in 3-4-1 model (see Refs.\cite{Kelso:2013zfa,Kelso:2014qka} for studies in 3-3-1 models). Our goal is to compute all corrections to the muon magnetic moment using the Public g-2 Code \cite{Queiroz:2014zfa} and determine whether or not this model is a plausible framework to accommodate the muon magnetic moment excess.
This paper is divided as follows: in the section II we describe the model, whereas in section III we discuss existing bounds. We review the muon magnetic moment and present our main results in section IV. Lastly in section V we draw our conclusions.

\section{3-4-1 Model}

There are several 3-4-1 models that lead to different fermion and gauge boson content \cite{Dias:2013kma,Palcu:2009ks,Palcu:2009ky,Palcu:2009kb,Riazuddin:2008yx}. Those models are based on the most general expression for the electric charge operator for the  $SU(4)_L\otimes U(1)_X$ found to be,
\begin{equation}\label{ch} 
Q=aT_{3L}+\frac{b}{\sqrt{3}}T_{8L}+
\frac{c}{\sqrt{6}}T_{15L}+ XI_4, \end{equation} 
where $a,b$ and $c$ are free parameters that set the fermion and scalar multiplets as well as the the gauge boson content. $T_{iL}=\lambda_{iL}/2$, with $\lambda_{iL}$ the Gell-Mann matrices for $SU(4)_L$ normalized as Tr$(\lambda_i\lambda_j)=2\delta_{ij}$, and $I_4=Dg(1,1,1,1)$ is the diagonal $4\times 4$ unit matrix. In this work, we focus on the $b=c=1$ setting, whose the covariant derivative is found to be $iD^\mu=i\partial^\mu -g_4 \lambda_{L\alpha} A^\mu_\alpha/2-g_XXB^\mu$, where

\begin{equation}
\frac{1}{2}\lambda_{L\alpha} A^\alpha_\mu=\frac{1}{\sqrt{2}}\left(
\begin{array}{cccc}D^0_{1\mu} & W^{+}_\mu & K^{+}_\mu & X^{+}_\mu\\ 
W^{-}_\mu & D^{0}_{2\mu} &  K^{0}_{\mu} &  X^{0}_\mu\\
K^{-}_\mu & K^{\prime 0}_{\mu} & D^{0}_{3\mu} & Y^{0}_\mu\\
X^{-}_\mu & X^{\prime 0}_\mu & Y^{\prime 0}_\mu & D^{0}_{4\mu} \end{array}\right).
\label{Covderiv}
\end{equation}

The lepton representation in this model is:

\begin{equation}
\begin{array}{ccccc}
f_{\alpha L}=\left(\begin{array}{c}
\nu_{\alpha}\\
e_{\alpha}\\
E_{\alpha}^-\\
E_{\alpha}^{'-}\end{array}\right)_{L}\sim(1,4,-3/4) &  &  &  & e_{\alpha L}^{c}\sim(1,1,1),  E_{\alpha L}^{c}\sim(1,1,1), E_{\alpha L}^{'c}\sim(1,1,1), \end{array}
\label{Eq.18}
\end{equation}

where $\alpha = 1,2,3$ are family indexes. In order to cancel all the quirial anomalies, two left handed quark families must transform as 4-plets and the other one as an anti 4-plet

\begin{equation}
\begin{array}{ccc}
Q_{iL}=\left(\begin{array}{c}
d_{i}^{\prime}\\
u_{i}\\
U_{i}\\
U_{i}^{\prime}\end{array}\right)_{L}\sim(3,4^{\star},5/12) &  & Q_{3L}=\left(\begin{array}{c}
u_3\\
d_3\\
D_{3}\\
D_{3}^{\prime}\end{array}\right)_{L}\sim(3,4,-1/12)\end{array}\label{Eq.19}\end{equation}
\begin{equation}
\begin{array}{c}
(d^{c}_{3L}),(d^{c}_{iL}),(D^{c}_{iL}),(D^{\prime c}_{iL})\sim(3^{\star},1,+1/3);
(u^{c}_{3L}),(u^{c}_{iL}),(U^{c}_{L}),(U^{\prime c}_{L})\sim(3^{\star},1,-2/3)
\end{array}\label{Eq.20}\end{equation} where $i=1,2$ and $\alpha = 1,2,3$ are family indexes. From Eqs.(\ref{Covderiv})-(\ref{Eq.20}) we see the 3-4-1 particle spectrum is rather rich.  As far as the muon magnetic moment is concerned few particles will be relevant as we shall see further. We now briefly discuss the scalar sector.

\subsection{\label{sec:sub3b}Scalar Content}
To generate masses for the fermions and gauge bosons of the mode the following multiplets are needed,
\begin{eqnarray}\nonumber 
\langle\phi^T_1\rangle&=&\langle(\phi^0_1,\phi^+_1,\phi^{\prime +}_1,\phi^{\prime\prime +}_1)\rangle=(v,0,0,0)\sim[1,4^*,3/4], \\ \nonumber
\langle\phi^T_2\rangle&=&\langle(\phi^-_2,\phi^0_2,\phi^{\prime 0}_2,\phi^{\prime\prime 0}_2)\rangle=(0,v',0,0)\sim[1,4^*,-1/4], \\ \nonumber
\langle\phi^T_3\rangle&=&\langle(\phi^-_3,\phi^0_3,\phi^{\prime 0}_3,\phi^{\prime\prime 0}_3)\rangle=(0,0,V,0)\sim[1,4^*,-1/4], \\ \nonumber
\langle\phi^T_4\rangle&=&\langle(\phi^-_4,\phi^0_4,\phi^{\prime 0}_4,\phi^{\prime\prime 0}_4)\rangle=(0,0,0,V')\sim[1,4^*,-1/4], \\ \label{scalars}
\end{eqnarray}
where we assume the hierarchy $V\sim V^\prime >> v\sim v^{\prime} \simeq 246$~GeV. After the spontaneous symmetry breaking mechanism we find the gauge coupling to be related as follows,

\begin{eqnarray}
g_4 = g, \qquad&\mbox{and}&\qquad\frac{1}{g^{\prime 2}}=\frac{1}{g^2_X}+\frac{1}{2g^2}, \label{match}
\end{eqnarray}where $g$ and $g^\prime$ are the gauge coupling constants of the $SU(2)_L$ and $U(1)_Y$ gauge groups of the SM, respectively.

\subsection{\label{sec:sub3c}Gauge boson masses and currents}

Using the configuration of the vev described in Eq.(\ref{scalars}) we find the masses of the gauge bosons to be,  
\begin{eqnarray}\nonumber
M^2_{W^\pm}&=&\frac{g^2}{4}(v^2+v^{\prime 2}), \quad M^2_{K^\pm}=\frac{g_4^2}{4}(v^2+V^2), \\ \nonumber
M^2_{X^\pm}&=&\frac{g^2}{4}(v^2+V^{\prime 2}), \quad
M^2_{K^0(K^{\prime 0})}=\frac{g^2}{4}(v^{\prime 2}+V^2), \\ \nonumber
M^2_{X^0(X^{\prime 0})}&=&\frac{g^2}{4}(v^{\prime 2}+V^{\prime 2}), \quad M^2_{Y^0(Y^{\prime 0})}=\frac{g^2}{4}(V^2+V^{\prime 2}). \\ \label{chmass}
\end{eqnarray}

For the neutral gauge bosons the $4\times4$ mass matrix has a zero eigenvalue corresponding to the photon. For the remainder $3\times3$ matrix we obtain the mass eigenvectors $Z_{\mu}$, $Z_{\mu}^{\prime}$ and $Z_{\mu}^{\prime \prime}$. In the approximation $V=V^{\prime}$ the field $Z_{\mu}^{\prime\prime}=A_{\mu}^{8}/\sqrt{3}-\sqrt{2/3}A_{\mu}^{15}$ decouples from the other two and acquires a mass $M_{Z_{\mu}^{\prime \prime}}^{2}=g^{2}V^{2}$. The latter is not relevant to the muon magnetic moment so it will be hereafter ignored. The other two massive gauge bosons are mixed and have a mass matrix in the basis $Z-Z^{\prime}$ given by, 
\begin{equation}
\frac{g_4^2}{C^2_W}\left(\begin{array}{cc} v^2 & \sqrt{2}\delta v^2 S_W \\
\sqrt{2}\delta v^2 S_W & \frac{2 \delta^2}{S^2_W}[v^2(S^4_W+C^4_W)+V^2 C^4_W]\end{array}\right),
\end{equation}where $\delta=g_X/(2g)$, and $S_W$ and $C_W$ are the sine and cosine of the electroweak mixing angle. 

By diagonalizing this mass matrix we get the two physical neutral gauge bosons
\begin{eqnarray}\nonumber
Z_1^\mu&=&Z^\mu \cos\theta+Z^{\prime\mu} \sin\theta \; ,\\ \label{mixing}
Z_2^\mu&=&-Z^\mu \sin\theta+Z^{\prime\mu} \cos\theta, 
\end{eqnarray} 
where the mixing angle is given by
\begin{equation} \label{tan} \tan(2\theta) = \frac{2 \sqrt{2} \delta v^2 S^3_W}
{2 \delta^2[v^2(S^4_W+C^4_W)+V^2 C^4_W]-v^2 S^2_W}. 
\end{equation}
\noindent  

We will take the limit which the mixing angle is small enough so that $Z_1\equiv Z$ and $Z_2\equiv Z^{\prime}$. Now we have shown mass terms of the gauge bosons we present below the charged and neutral currents relevant for the muon magnetic moment, 

\begin{equation}
{\cal L}^{CC} \supset -\frac{g}{\sqrt{2}} \left( \overline{l_L} \gamma^{\mu} \frac{(1-\gamma^5)}{2} E K^0_{\mu} + \overline{l_L} \gamma^{\mu} \frac{(1-\gamma^5)}{2} E^{\prime} X^0_{\mu}\right),
\label{CC}
\end{equation}

\begin{equation}
{\cal L}^{NC} \supset \frac{g}{2C_W} \bar{l} \gamma^{\mu} \left( g_V - g_A \gamma^5\right)l Z^{\prime}
\label{NC}
\end{equation}where

\begin{equation}
g_V = \frac{1/2 +S_W^2}{\sqrt{2-3S_W^2}}; g_A= \frac{C_{2W}}{2\sqrt{2-3S_W^2}}.
\end{equation}

We point out that Eq.\ref{CC} and Eq.\ref{NC} which refer to the charged and neutral current contain only the relevant interactions for the muon magnetic moment. Obviously the complete charged and neutral currents are comprised of many other interactions involving all fermions and gauge bosons of the model. We decided not to dwell on unnecessary details and for this reason we present only the terms in Eq.(\ref{CC}) and Eq.(\ref{NC}).

\section{\label{sec:sec4}Existing Bounds}

Searches for resonances on the $pp\rightarrow l^+l^-+X$ reaction with CMS data resulted on the exclusion of $M_Z^{\prime}$ below of $~$ 2.5TeV \cite{Coutinho:2013lta}. However the $Z^{\prime}$ gauge boson in this study does not couple to the fermions like ours, so it is unclear what would be the limit in our model. If we are conservative and take this limit on the $Z^{\prime}$  mass to lie in the TeV scale, it can be translated into a lower limit of $3$TeV on the scale of symmetry breaking of the model. Moreover, using precise measurements on SM Z boson a lower bound of $2$TeV has been placed on the $Z^{\prime}$ mass \cite{Sanchez:2008qv} precisely in this model. The latter implies into a scale of symmetry breaking greater than 5.7 TeV. Furthermore flavor changing neutral current studies rule out $Z^{\prime}$ masses below 11 TeV, depending on the parametrization scheme used in the quark sector \cite{Sanchez:2008qv}. In summary, there are several bounds possibly applicable to this model, but the limit of $5.7$~TeV on the scale of symmetry breaking of the model is definitely the most robust and will be our reference from now on. After discussion the existing bounds, we review the muon magnetic moment and show our results.

\section{Results}

The Dirac equation predicts a muon magnetic moment 

\begin{equation}
\overrightarrow{M}_{\mu} = g_{\mu} \left( \frac{e}{2m_{\mu}} \right) \overrightarrow{S}
\end{equation}

with gyromagnetic ratio $g_{\mu}=2$. However, quantum loop effects lead to a small calculable deviation from $g_{\mu}=2$, the anomalous magnetic moment, parametrized by $a_{\mu} = (g_{\mu}-2)/2$. The SM prediction for the $a_{\mu}$ is generally divided into three parts: electromagnetic (QED), electroweak (EW) and hadronic contributions \cite{PDG}. The QED part, which is by far the dominant contribution in the SM, includes all photonic and leptonic ($e,\mu,\tau$) contributions and has been computed up to four loops and estimated at the 5 loops\cite{Aoyama:2007dv,Kinoshita:2005sm,Hanneke:2008tm}. The EW contribution comprises $W^{\pm},Z$ and Higgs bosons, and has been calculated up to three loops. The hadronic contributions are the most uncertain though and can not be calculated by first principles. The hadronic vacuum polarization is determined from $e^+e^- \rightarrow$ hadrons or $\tau \rightarrow$ hadrons data \cite{PDG}. The next largest uncertainty is associated with hadronic light-by-light scattering, which is computed using hadronic models that correctly reproduce the properties of QCD \cite{Boughezal:2011vw}. In summary, the SM prediction for $a_{\mu}$ is \cite{carey},

\begin{equation}
a_{\mu}^{SM} = (116591785 \pm 51) \times 10^{-11}.
\end{equation}

The E821 experiment at Brookhaven National Lab has reported \cite{E821_2,E821_3},

\begin{equation}
a_{\mu}^{E821}= (116592080 \pm 63) \times 10^{-11}.
\end{equation}
Thus, 

\begin{equation}
\Delta a_{\mu} (E821 -SM) = (295 \pm 81) \times 10^{-11},
\label{deltaa}
\end{equation}which points to a $3.6\sigma$ discrepancy. The current theoretical error of $\pm 51 \times 10^{-11}$ is driven by the $\pm 39 \times 10^{-11}$ uncertainty on lowest-order hadronic contribution and the $\pm 26 \times 10^{-11}$ uncertainty on the hadronic light-by-light contribution as aforesaid\cite{carey}. A long standing effort predicts that the uncertainty on the lowest-order hadronic contribution could be dwindled to $25 \times 10^{-11}$ with existing data and further work on the hadronic light-by-light contributions could reduce the total SM error to possibly $\pm 30 \times 10^{-11}$ \cite{carey}. The Fermilab experiment will play an important role in this setup with the proposed experimental error of $\pm 16 \times 10^{-11}$. The combined effort in theory and experiment front is expect to reach

\begin{equation}
\Delta a_{\mu} ({\rm Fermilab} -SM) = (295 \pm 34) \times 10^{-11}.
\label{deltaa}
\end{equation}

\begin{figure}[!t]
\centering
\includegraphics[scale=0.6]{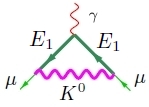}
\includegraphics[scale=0.6]{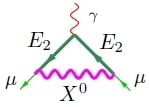}
\includegraphics[scale=0.6]{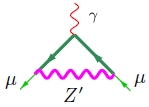}
\caption{Feynmann diagrams that contribute to the muon magnetic moment. The first two involve exotic charged leptons and charged gauge bosons, and last one the $Z^{\prime}$ gauge boson.}
\label{feyn2}
\end{figure}

This discrepancy in Eq.(\ref{deltaa}) will be referred as future sensitivity for the muon magnetic moment. That being said we computed all corrections to the muon magnetic moment stemming from our model using the public code in Ref.\cite{Queiroz:2014zfa}. In this model those contributions arise from the presence of exotic charged leptons and the neutral vector boson $(Z^{\prime})$ (See Fig \ref{feyn2}). In this work we ignore the contributions stemming from charged and neutral scalars since they are suppressed by the muon mass.

\begin{figure}[!h]
\centering
\includegraphics[scale=0.8]{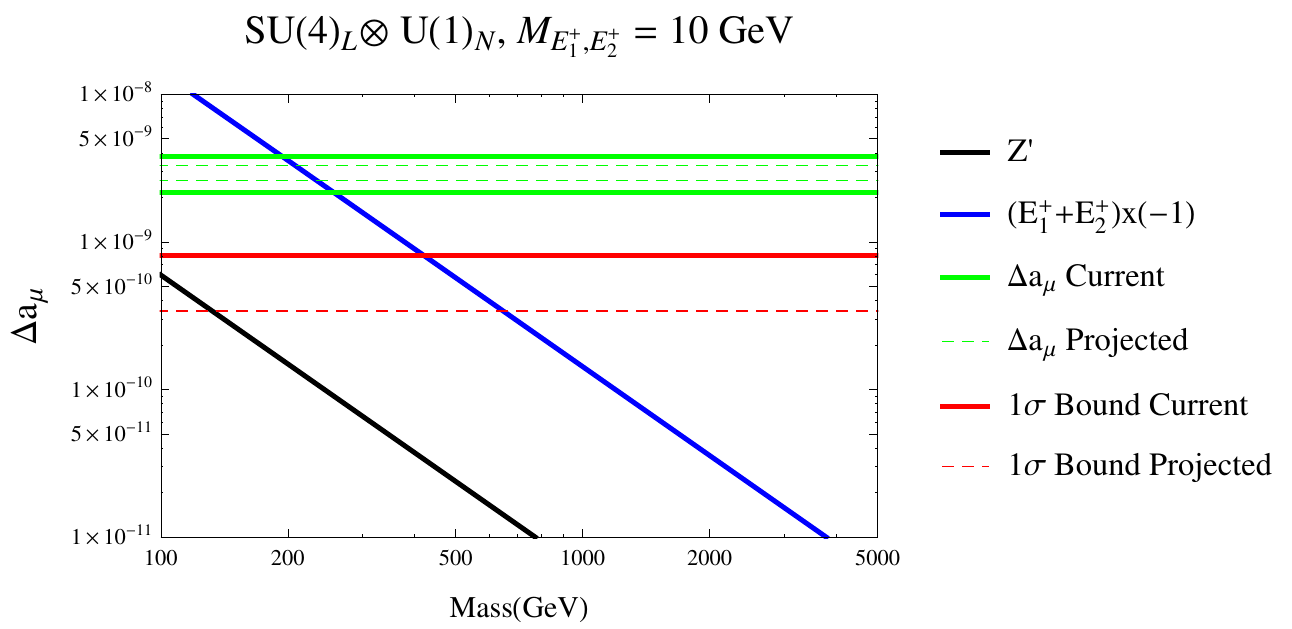}
\label{fig2}
\caption{Individual corrections to the muon magnetic moment as function of the $K^0$, $X^0$ and $Z^{\prime}$ masses for $M_{E_1,E_2}=10$~GeV.}
\includegraphics[scale=0.8]{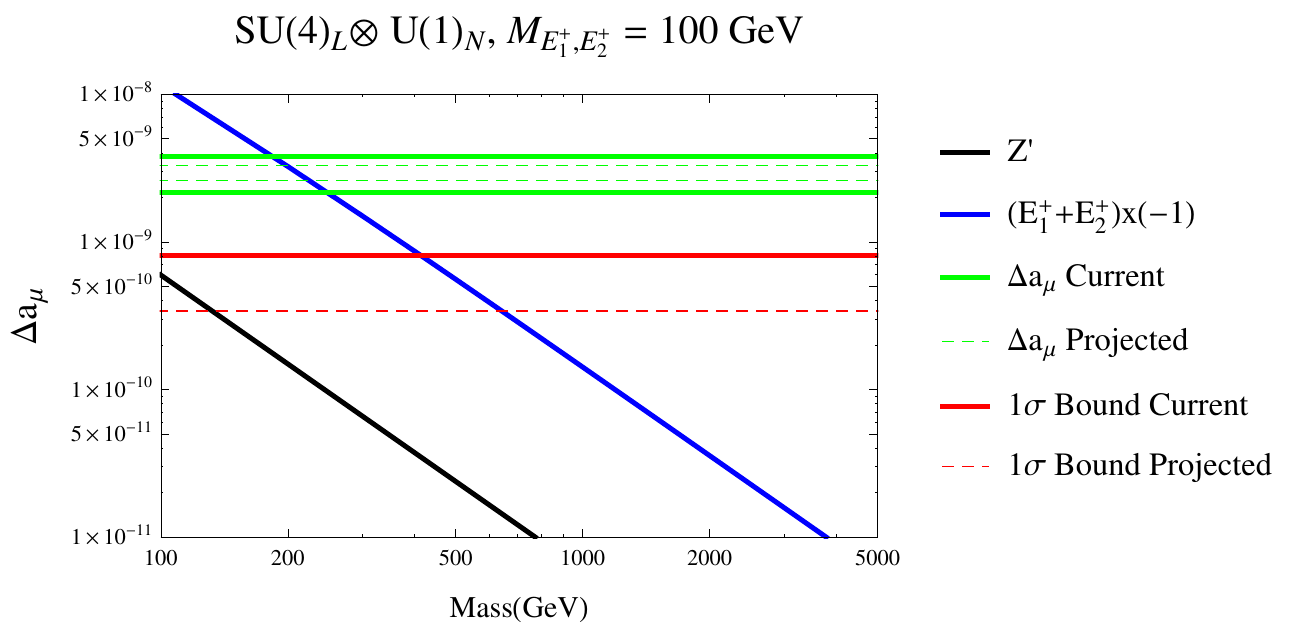}
\label{fig3}
\caption{Individual corrections to the muon magnetic moment as function of the $K^0$, $X^0$ and $Z^{\prime}$ masses for $M_{E_1,E_2}=100$~GeV.}
\includegraphics[scale=0.8]{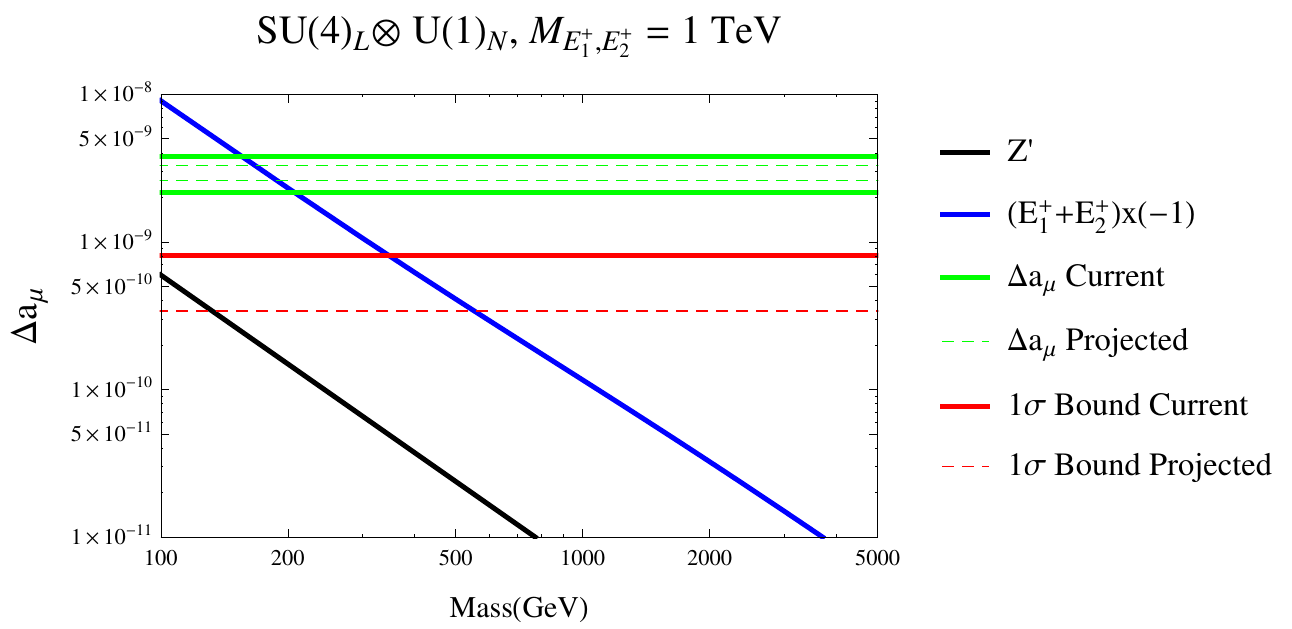}
\label{fig4}
\caption{Individual corrections to the muon magnetic moment as function of the $K^0$, $X^0$ and $Z^{\prime}$ masses for $M_{E_1,E_2}=1$~TeV.}
\end{figure}

\begin{figure}[!h]
\centering
\includegraphics[scale=0.8]{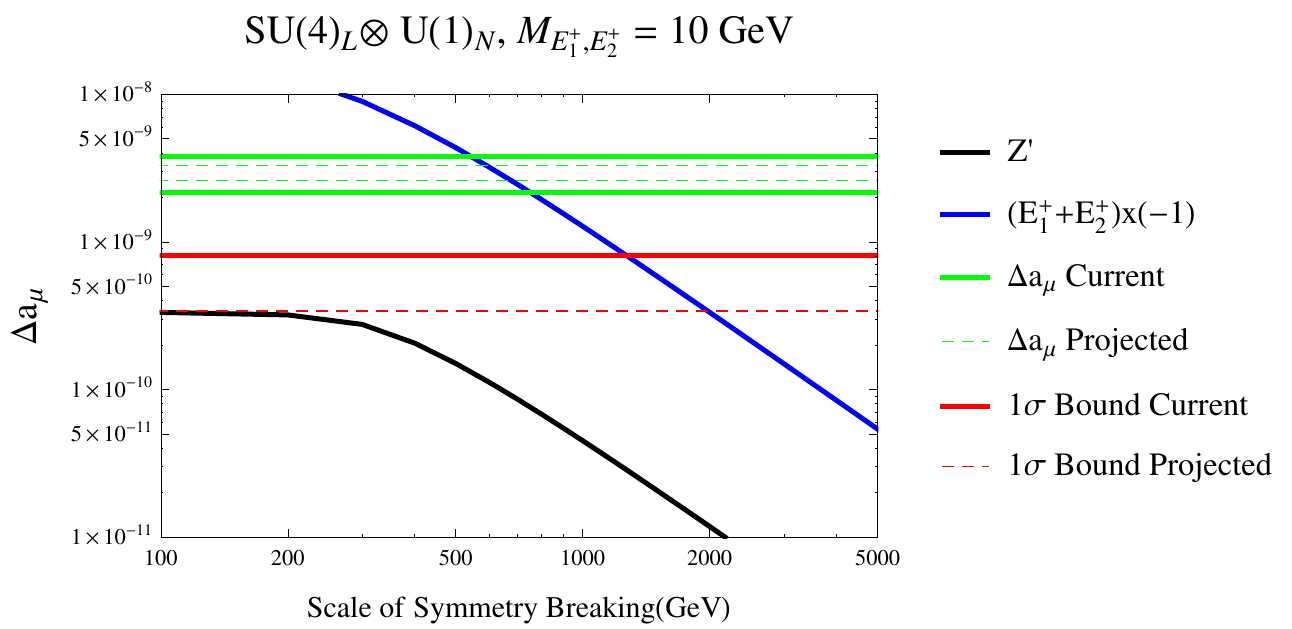}
\label{fig5}
\caption{Individual corrections to the muon magnetic moment stemming in terms of the scale of symmetry breaking for $M_{E_1,E_2}=10$~GeV.}
\includegraphics[scale=0.8]{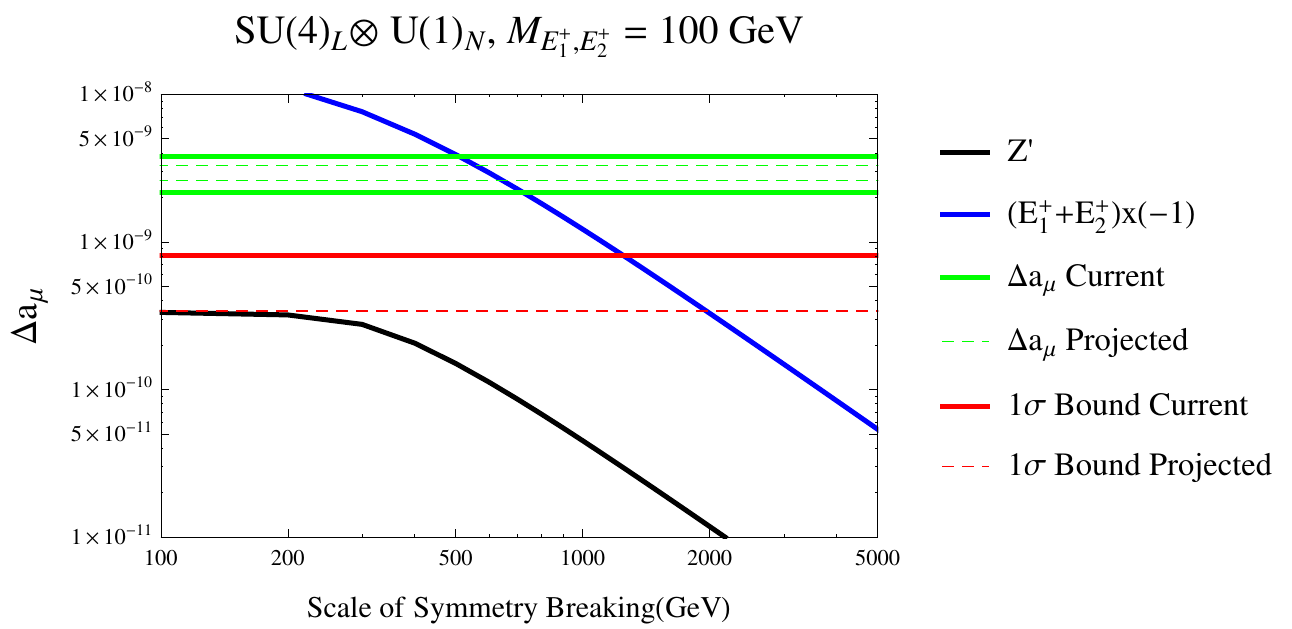}
\label{fig6}
\caption{Individual corrections to the muon magnetic moment stemming in terms of the scale of symmetry breaking for $M_{E_1,E_2}=100$~GeV.}
\includegraphics[scale=0.8]{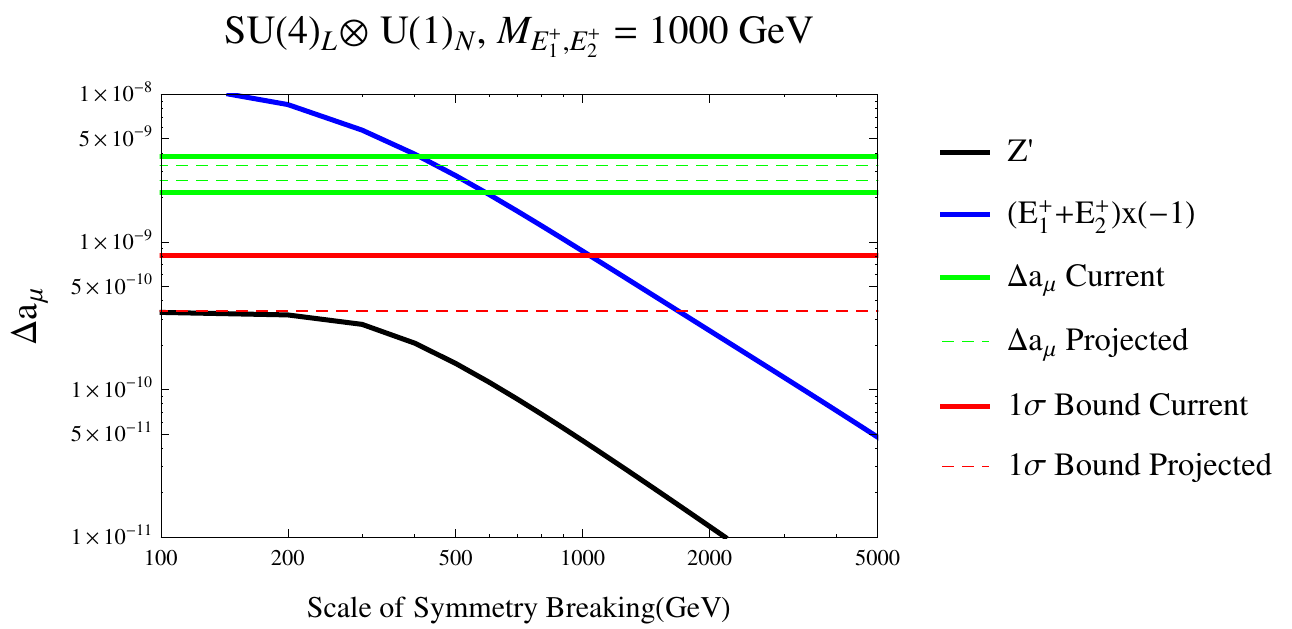}
\label{fig7}
\caption{Individual corrections to the muon magnetic moment stemming in terms of the scale of symmetry breaking for $M_{E_1,E_2}=1$~TeV.}
\end{figure}

\begin{figure}[!h]
\centering
\includegraphics[scale=0.8]{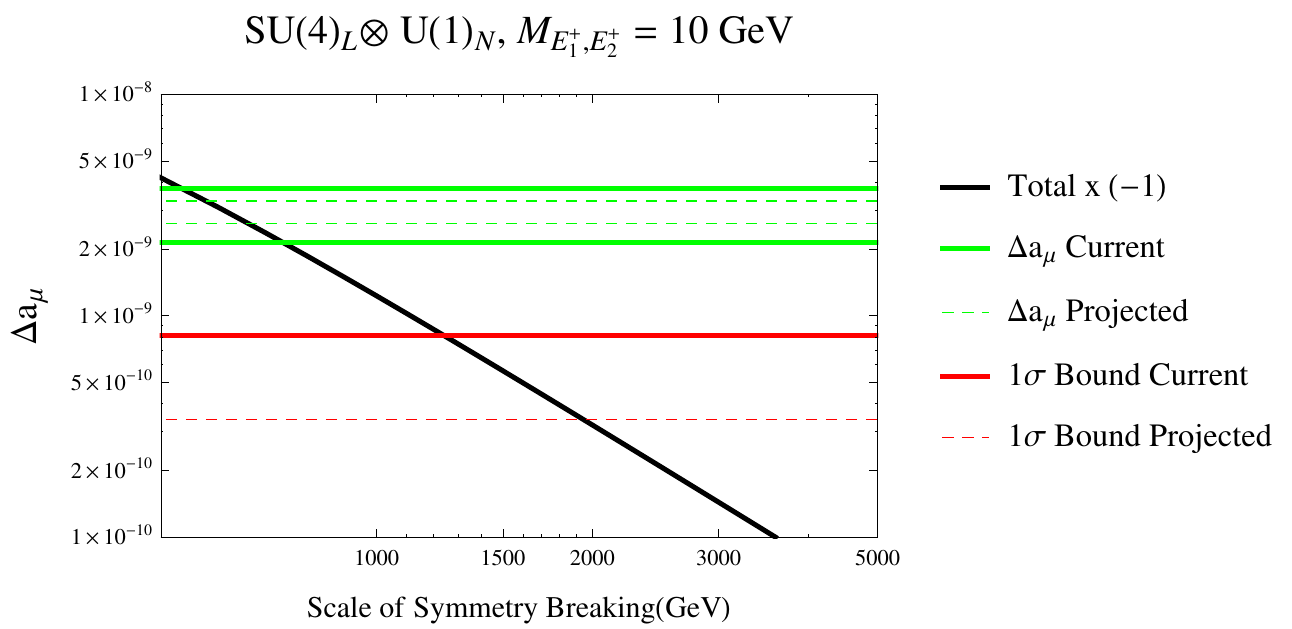}
\label{fig8}
\caption{The total contribution to the muon magnetic moment coming as function of the scale of symmetry breaking for $M_{E_1,E_2}=10$~GeV. }
\includegraphics[scale=0.8]{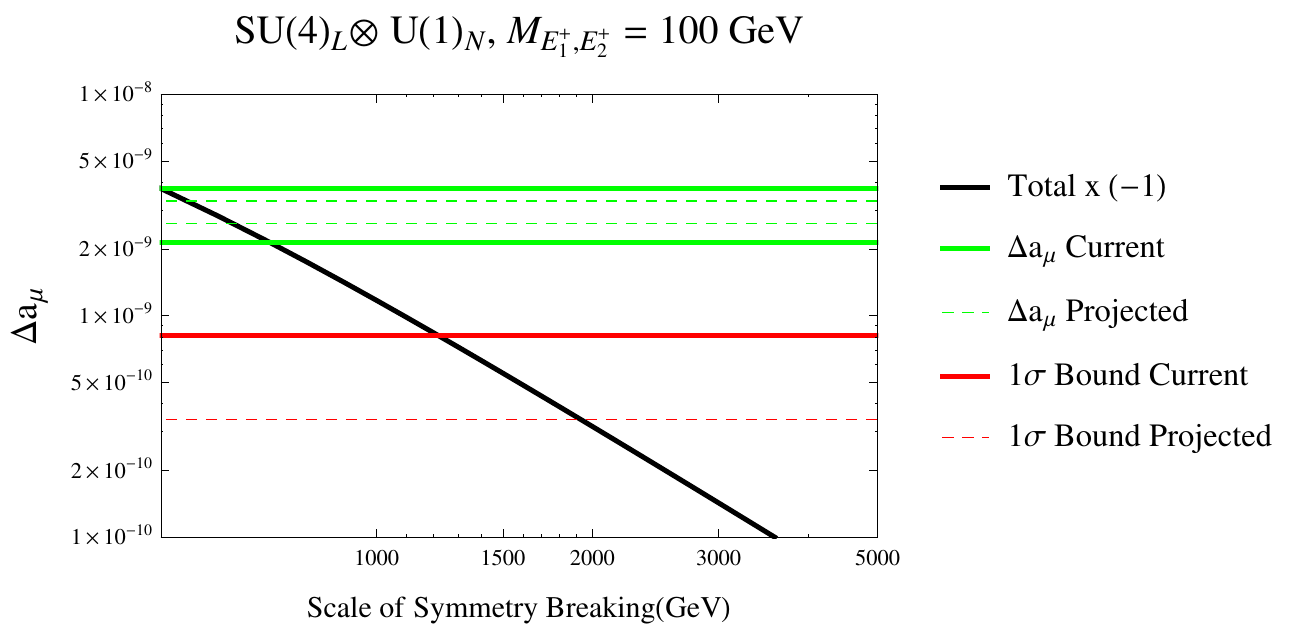}
\label{fig9}
\caption{The total contribution to the muon magnetic moment coming as function of the scale of symmetry breaking for $M_{E_1,E_2}=100$~GeV.}
\includegraphics[scale=0.8]{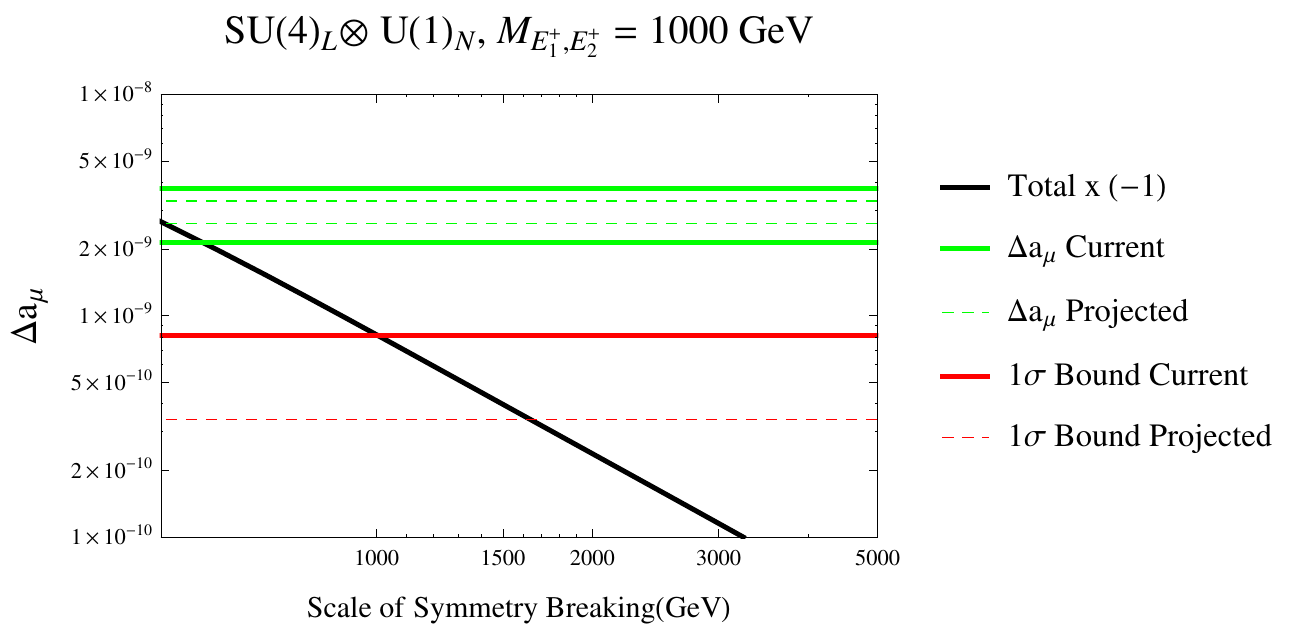}
\label{fig10}
\caption{The total contribution to the muon magnetic moment coming as function of the scale of symmetry breaking for $M_{E_1,E_2}=1$~TeV. }
\end{figure}

In Figs.2-4 we exhibit the individual contributions of each one of those particles as a function of the neutral ($Z^{\prime}$) and the gauge bosons $K^0$ and $X^0$ masses for heavy charged fermion masses of $10$~GeV,$100$~GeV and $1$~TeV respectively. We notice that besides giving a negative correction to the muon magnetic moment, the charged fermions give rise to the most sizeable contribution for $M_{E,E^{\prime}}=10,100,1000$~GeV. In all figures we have multiplied the charged fermion contribution by minus one simply to show them in the plots where we use LogLog scale. The $Z^{\prime}$ correction despite being positive is rather small and thus negligible. For $M_{E,E^{\prime}}=10,100$~GeV we find current and projected $1\sigma$ bounds of 400 GeV and 700 GeV respectively on the mass of the gauge bosons so that their contributions do not exceed the error bars. For $M_{E,E^{\prime}}=1$~TeV we obtain a current and projected $1\sigma$ bounds of 350 GeV and 600 GeV on the masses of the bosons $K^0$ and $X^0$. We emphasize that those are the strongest constraints on the masses of those gauge bosons in the literature. 

In Figs.5-7, for completeness we show the individual contributions as a function of the scale of symmetry breaking. In Figs.8-10 we exhibit the total contributions to the muon magnetic moment as a function of the scale of symmetry breaking. We conclude that for $M_{E,E^{\prime}}=10,100$~GeV scales of symmetry breaking smaller than $1.2$~TeV and $2$~TeV are excluded using current and projected bounds respectively. Moreover, for $M_{E,E^{\prime}}=1$~TeV, using current and projected $1\sigma$ bounds we then rule out scales smaller than $1$~TeV and $1.6$~TeV respectively. Thus our study of the muon magnetic moment provides robust and complementarity bounds on the mass spectrum and scale of symmetry breaking of the model. We emphasize that our field content is quite different from 3-3-1 \cite{Dong:2014wsa,Alves:2012yp,Alves:2011kc,
Caetano:2012qc,Pires:2014xsa,Queiroz:2010rj} and 3-4-1 models \cite{Dias:2013kma,Palcu:2009ks,Palcu:2009ky,Palcu:2009kb,Riazuddin:2008yx} proposed in the literature, therefore a new look into the muon magnetic moment contributions stemming from this 3-4-1 model was necessary.

\section{Conclusions}

The muon magnetic moment has been addressed in the context of 3-4-1 models, but in different versions. In this work we investigate g-2 in a 3-4-1 comprised of exotic charged leptons. The presence of exotic charged leptons induce negative and sizable corrections to the muon magnetic moment. Since the overall contribution to g-2 is negative, we place stronger bounds on the mass and scale of symmetry breaking of the model by enforcing the total correction to lie within the current and projected error bars. 

We computed the contributions for exotic charged lepton masses of $M_{E,E^{\prime}}=10,100,1000$~GeV as shown in Figs.2-7. For $M_{E,E^{\prime}}=10,100$~GeV we obtain current and projected $1\sigma$ bounds of 400 GeV and 700 GeV respectively on the mass of the neutral gauge bosons ($K^0$ and $X^0$). For $M_{E,E^{\prime}}=1$~TeV we find the upper bounds $M_{K^0,X^0} \geq 350$~GeV $M_{K^0,X^0} \geq 600$GeV, using current and projected sensitive respectively. We emphasize that those are the most stringent limits on the masses of those gauge bosons in the literature. 

To summarize, we conclude that the inclusion of exotic charged leptons in the spectrum of 3-4-1 models, drastically chances the bounds previously derived in the literature on scale of symmetry breaking of 3-4-1 models based on the muon magnetic moment, showing the importance of our work, which has discussed and derived the impact on the g-2 stemming from exotic charged leptons. Despite our reasoning be focused on 3-4-1 models, our results are somewhat general and applicable to any particle physics model that evokes exotic charged leptons with Lagrangians similarly presented here.

\section*{ACKNOWLEDGMENTS}

DC is partly supported by the Brazilian National Council for Scientific and Technological Development (CNPq) Grant 484157/2013-2.


\end{document}